\documentclass[preprint,5p,times,twocolumn]{elsarticle}

\usepackage{epsfig}

\usepackage{amssymb}
\usepackage{amsthm}
\usepackage{array} 
\newcolumntype{N}{c} 
\usepackage{amsmath}
\usepackage{placeins} 
\usepackage{booktabs} 
\usepackage{nicematrix}
\usepackage{xcolor}
\usepackage{verbatim}
\usepackage{caption}

\usepackage{pdfpages} 
\usepackage{hyperref}
\usepackage{comment}
\usepackage{multirow}

\usepackage{lineno}

\makeatletter

\newcommand{\volumedash}{%
  \makebox[0pt][l]{%
    \ooalign{\hfil\hphantom{$\m@th V$}\hfil\cr\kern0.08em--\hfil\cr}%
  }%
}
\makeatother

\journal{Soft Matter}

\begin{document}

\begin{frontmatter}




\title{Rheological and Photoelastic Response of Hydrated Soft Granular Particles}


\author[inst1,corr_author]{Brandon Hayes}
\author[inst1]{Krishnaroop Chaudhuri}
\author[inst1]{Rylan Hodgson}
\author[inst2]{Benjamin McMillan}
\author[inst1]{Ruby Gans}
\author[inst1]{Hale Burke}
\author[inst3]{Stephanie McNamara}
\author[inst4]{Thomas Chalklen} 
\author[inst1]{Nathalie M. Vriend}

\affiliation[inst1]{organization={University of Colorado Boulder},
            addressline={Paul M. Rady Department of Mechanical Engineering}, 
            city={Boulder},
            state={Colorado},
            postcode={80309}, 
            country={USA}}
            
\affiliation[inst2]{organization={University of Bristol},
            addressline={School of Engineering Mathematics and Technology, Ada Lovelace Building}, 
            city={University Walk},
            state={Bristol},
            postcode={BS8 1TW}, 
            country={United Kingdom}}  
            
\affiliation[inst3]{organization={University of Colorado Boulder},
            addressline={Department of Physics}, 
            city={Boulder},
            state={Colorado},
            postcode={80309}, 
            country={USA}}            

\affiliation[inst4]{organization={University of Cambridge},
            addressline={Department of Earth Sciences}, 
            city={Downing Street},
            state={Cambridge},
            postcode={CB2 3EQ}, 
            country={United Kingdom}}              

\fntext[corr_author]{Corresponding Author: brandon.hayes@colorado.edu}
        

\begin{abstract}
Photoelasticity is a qualitative and quantitative optical technique to image internal stress distributions in transparent materials. In the past few decades, discrete photoelastic particles have been used as a proxy for dry granular materials in both static, quasistatic, and dynamic analogue experiments. The technique allows the visualization of force chains, determination of the location and magnitude of contact forces, and outputs a stress tensor for each particle with shear and normal stress components. To date, little to no work has investigated photoelastic suspensions, where photoelastic granular particles are immersed in a fluid medium, despite its relevance in industrial and natural applications. The introduction of a fluid phase yields additional considerations in the rheological and photoelastic behavior of our proxy particles. In this manuscript, we summarize the state-of-the-art in resolving forces in immersed photoelastic granular materials. We introduce characterization techniques to probe changes in rheological and optical properties of hydrated photoelastic particles, and we report considerations for use of photoelastic particles in immersion-based experiments. We intend for this work to provide the leading framework to study the hydrodynamic interactions in 2D systems of photoelastic particles immersed in a fluid medium. 
\end{abstract}


\end{frontmatter}

\section{Introduction}

Flows of granular material sculpt the natural world and underpin countless industrial processes. While much progress has been made in understanding the rheological properties of these dynamic particle-based systems \cite{jenkins_theory_1983, GDRMiDi2004}, a complete mathematical description of their behavior remains elusive. The large number of grains and the complex physics governing their interactions pose significant challenges in modeling, numerically simulating, and experimenting with granular flows \cite{dauxois_confronting_2021}. In particular, the inherently non-uniform distribution of stress within granular materials is central to understanding their behavior \cite{jaeger_physics_1992}, yet measuring interparticle forces from experiments is often not possible.
\\
\\
In the simplest case of dry cohesion-free grains, photoelastic techniques have been used to probe force networks in quasi-two-dimensional granular material. Pioneering work by Behringer \textit{et al.} transformed granular photoelasticity from a qualitative method \cite{Wakabayashi1950,Drescher1972} to a quantitative and powerful experimental tool \cite{Jaeger1996, howell_1999, behringernature}. Since then, the technique has provided insight into a wide range of granular phenomena \cite{Bares2017,Majmudar2007,Clark2012,owens_sound_2011,mahabadi_impact_2017,Tang2018,abed_zadeh_enlightening_2019}. Much modern photoelastic research relies on the inverse methods first introduced by Daniels \textit{et al.} \cite{pegspaper}, which have been shown to accurately quantify both normal and frictional forces acting on photoelastic particles \cite{PEGS_Correction_McMillan_2025}. These methods have recently been extended to account for inertial effects and deformed contact regions \cite{Fossil_Photoelasticity_McMillan2026}, further broadening the applicability of granular photoelasticity to dynamic and highly confined systems.
\\
\\
Although photoelasticity has been used to study the rheology of dry granular material \cite{thomas_2019a,thomas_2019b,poon_2023,Fazelpour2023}, many important granular flows in nature involve an interstitial fluid and their rheological behavior remains poorly understood. Particle-particle forces often dominate in these flows, yet, hydrodynamic stress can play a non-trivial role \cite{guazzelli_rheology_2018}. Despite the widespread use of photoelasticity in dry granular systems, its application to granular suspensions is largely unexplored. The experimental data unlocked by photoelastic techniques will likely provide unique insight into the dynamics of these suspension flows; however, a careful characterization of how key properties of photoelastic materials change during immersion in a fluid has not yet been performed. This limitation presents a significant barrier to extracting accurate, reproducible, and reliable particle-scale photoelastic force measurements from fluid-saturated granular flows. 
\begin{table*}[t!]
\centering
\noindent
\hspace*{-0.75cm} 
\begin{tabular}{N N N N N}\toprule
\multicolumn{1}{N }{\textbf{}} & \multicolumn{3}{c }{\textbf{\textcolor{gray}{Mechanical Properties}}} & \multicolumn{1}{c }{\textbf{\textcolor{gray}{Optical Properties}}} \\  
\cmidrule(lr){2-4}
\cmidrule(ll){5-5}
\multicolumn{1}{ N }{\textbf{Material}} & \textbf{Shore-Hardness} & \textbf{Elastic Modulus ($MPa$)} & \textbf{Decay Time ($s$)} & \textbf{Stress-Optic Coefficient ($Br$)} \\ 
\cmidrule(lr){1-1}
\cmidrule(lr){2-4}
\cmidrule(ll){5-5}
\multicolumn{1}{ N }{Vishay (PSM-4)} & 60A & 4.00 & 0.001 & 3175 \\ 
\multicolumn{1}{ N }{Clear Flex 30} & 30A & 1.16 & 0.895 & 2877 \\ 
\multicolumn{1}{ N }{Clear Flex 50} & 50A & 2.91 & 0.020 & 2430 \\ 
\multicolumn{1}{ N }{\textcolor{blue}{WC-55}} & \textcolor{blue}{55A} & \textcolor{blue}{2.55} & \textcolor{blue}{0.112} & \textcolor{blue}{2121} \\ 
\multicolumn{1}{ N }{Acrylic} & 80D & 3200 & $\approx 0$ & 4.0 \\
\multicolumn{1}{ N }{Glass} & n/a & 70000 & $\approx 0$ & 2.43 \\
\bottomrule
\end{tabular}
\caption{Properties of Photoelastic Materials. Tabulates select mechanical and optical properties of common photoelastic materials. We define the decay time as the time required for the normalized tensile time-dependent stress relaxation modulus in the linear viscoelastic regime to decay to a value of 1/e. In this study, we use WC-55 which is highlighted in blue. The following material datasheets and the listed works are used to report properties for Vishay PhotoStress\texttrademark{}, Clear Flex 30, Clear Flex 50, acrylic, and glass when not experimentally measured \cite{pegspaper,stabile_clearflex50_reference,cramer2014quantification,vishay_photostress_coatings_2015,smoothon_clearflex5095_tb,bjb_wc55am_2025}. We note that Vishay PhotoStress\texttrademark{} materials have been discontinued and include it here for historical reference. 1 $Br$ = 10 $^{-12} \ m^2/N$}
\label{table: properties of materials}
\end{table*}
\\
\\
In this work, we investigate how hydration-induced swelling modifies the rheological and optical properties of a selected photoelastic material, specifically water-clear 55 (WC-55, BJB Materials, Tustin, CA, USA) polyurethane. WC-55 was selected due to its mechanical and optical properties as well as being a safer alternative to conventional Clear Flex 50. We consider photoelastic particles that have been submerged in water and quantify how mechanical properties such as the storage, loss, and elastic moduli are affected by swelling. In this paper, our findings show that the photoelastic response of the polymer is reduced by approximately $9\%$ and the elastic modulus is reduced by approximately 22$\%$ as particles are hydrated, and we identify key trends in the stress-optic coefficient response as a function of immersion time. We report that swollen photoelastic particles can be returned to near their initial pre-swelled states through annealing at $70 \ ^{\circ}C$. To our knowledge, this is the first study to systematically examine the mechanical and photoelastic effects of swelling in soft granular particles. These results provide practical guidelines for experiments with submerged photoelastic materials and establish a foundation for extending particle-scale force measurements to wet granular systems.


\section{Materials and Methods}

\subsection{Choice of Photoelastic Material}

We provide a list of common photoelastic materials in table \ref{table: properties of materials} and summarize their relevant mechanical and optical properties. The \textit{shore-hardness} measures surface resistance to indentation, and is determined with a durometer. The \textit{elastic modulus} measures the stiffness, or the material's resistance to elastic deformation under load, and is calculated from tensile testing or rheometer data. Harder materials typically feature a higher elastic modulus, but they measure different properties. The \textit{decay time} measures how fast the material relaxes, and is calculated as the time required for the normalized tensile time-dependent stress relaxation modulus to decay to a value of 1/e in the linear viscoelastic regime -- discussed in detail in section \ref{sec: rheological analysis}. Photoelastic calibration experiments, will which will be described in section \ref{sec: Photoelastic Calibration Experiment}, characterize the stress-optic coefficient of a given material. This material property quantifies the sensitivity of a photoelastic material, as it is a measure of how many fringes will result from a given prescribed load. 
\\
\\
To select the ``right'' material to use for photoelastic experiments, one should consider the \textit{applied force} and specific \textit{time scale} featured. Large forces may deform a material beyond linearity and can even lead to fracture, invalidating the assumptions of standard quantitative photoelastic analysis of rigid disks. In fast, dynamic experiments, relevant time scales are short and polymer materials can exhibit significant fossil photoelasticity \cite{Fossil_Photoelasticity_McMillan2026}, where the material stress relaxation time exceeds the relevant experimental times.
\\
\\
The ideal photoelastic material would be one that has infinite stiffness, instantaneous relaxation time, and a very high stress-optic coefficient. However, real materials have a finite stiffness, relaxation time, and stress-optic coefficient; therefore, the ``right'' photoelastic material is tuned and balanced to each individual experiment. To illustrate practical selection of the ``right'' photoelastic material, consider two hypothetical photoelastic experiments: (1) a slow, quasistatic flow in a shear-cell \cite{howell_1999,pegspaper,Carmeen_Lee_SFR_relationship} and (2) a fast, transient flow during an explosive impact \cite{wave_propagation_granular_1991_hard_disks_SHUKLA1991165}. In the first scenario, photoelastic disks are displaced at discrete strain intervals and generate forces up to $5 \ N$. From the criteria described above, polymer materials such as Clear Flex 50 or WC-55 would be suitable for this hypothetical experiment. Fossil photoelasticity is not a concern since the material stress relaxation time occurs significantly faster than the time scale of the experiment. In the second scenario, an explosive impact on photoelastic disks generates forces up to $2000 \ N$ in less than a millisecond. Here, stiffer materials such as acrylic, glass, or ceramics with near instantaneous relaxation times and a lower stress-optic coefficient would be suitable. The stress-optic coefficient is selected such that the maximum force loading on a material results in a sufficient number of fringes (we suggest $N < 5$) to visualize and analyze via quantitative photoelasticity.
\\
\\
We wish to stress \textbf{two health safety concerns} when working with cast polyurethane liquid materials: (1) nearly all polyurethane liquids contain sensitizers in their formulation that some users have or can develop allergic reactions to over time and (2) some polyurethane liquids, such as Clear Flex 50, contain an organomercury compound in their catalyst formulation which can be dangerous to the environment and human health. Care should be taken when working with these materials, proper ventilation should be used and personal protective equipment (PPE) should be worn.

\subsection{Fabrication of Soft Granular Particles}

\begin{figure*}[t!]
    \centering
    \includegraphics[width=\linewidth]{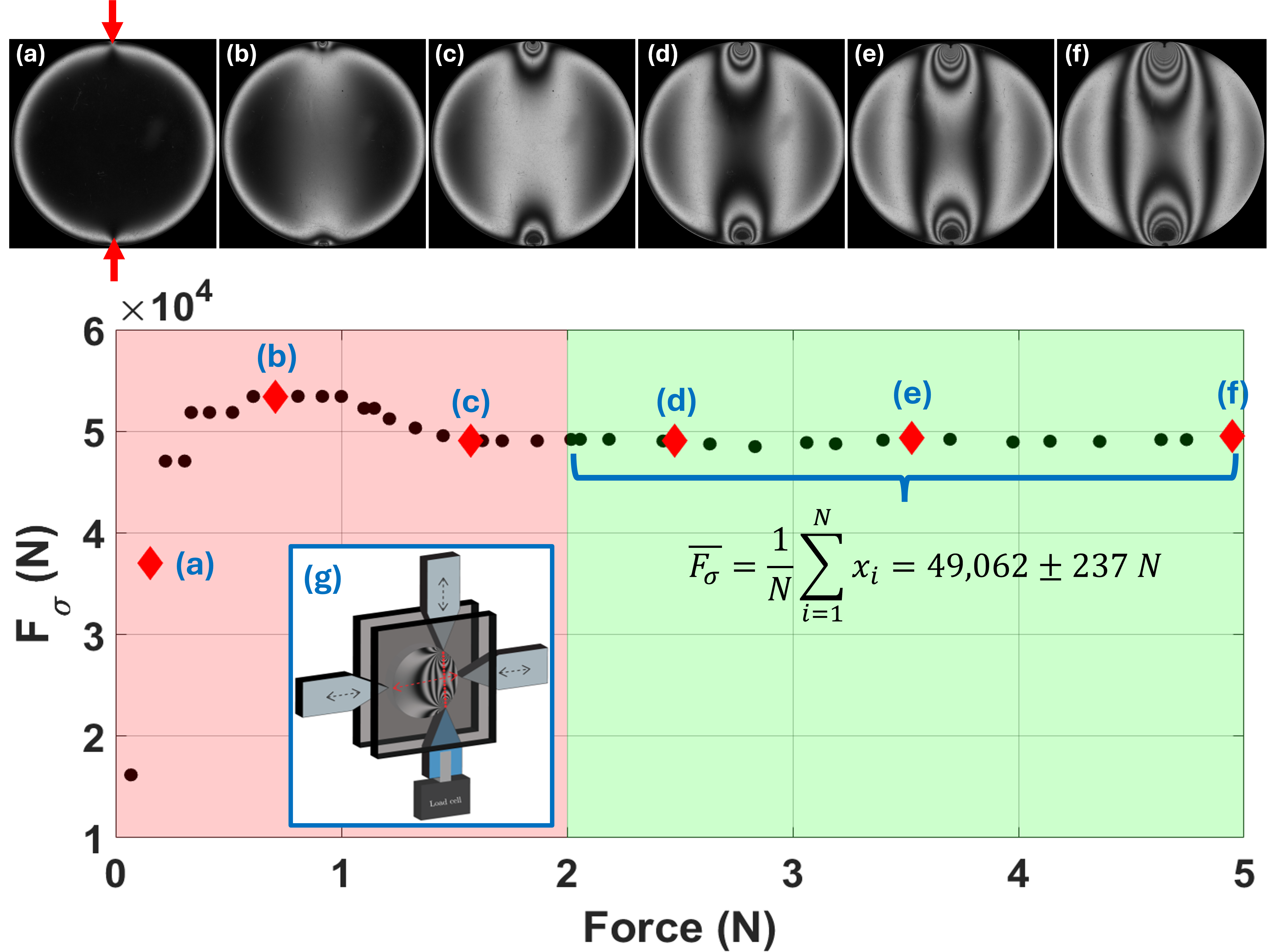}
    \caption{Photoelastic Calibration Experiment. The stress-optic coefficient is found using PEGS fringe inversion under diametric loading conditions. (a-f) Illustrate diametric loading for a $D = 22 \ mm$ and $H = 6 \ mm$ WC-55 particle. Below a critical force limit ($F = 2 \ N$ in this data), PEGS fringe inversion cannot be used to accurately estimate $F_\sigma$ due to insufficient number of fringes. Beyond the critical force limit, the value of $F_\sigma$ and thus the stress-optic coefficient, $F_\sigma = \lambda / Ch$, is theoretically constant. Here, $F_\sigma = 49,062 \ N$ which implies $C = 1980 \ Br$. (g) Shows the experimental setup used to apply diametric loading conditions, adapted with permission from McMillan et al. \cite{PEGS_Correction_McMillan_2025}}
    \label{fig: fringe fitting and test setup}
\end{figure*}

The goal of this work is to characterize swelling to inform future studies of fast, transient, \textit{wet} granular flows with expected forces ranging up to $5 \ N$. As such, in this study, we selected water-clear 55 (WC-55) polyurethane as our photoelastic material of choice due to satisfactory stiffness, relaxation time, and stress-optic coefficient for our planned experimental conditions. Photoelastic particles were casted out of commercial WC-55 polyurethane two-part resin and mixed in a 1A:2B mass ratio, per manufacturer's recommendation. We briefly summarize the casting process below but we refer readers to the methodology outlined by Cox \cite{Cox2016} and Bar\'es \cite{Bars2016ExperimentalOO} for detailed instructions. 
\\
\\
Primarily, the negative mold was fabricated out of aluminum containing cylindrical disks of diameter, D = 22 $mm$, and height, H = 6 $mm$. The surface was polished with Simichrome diamond polishing paste. To create the positive mold, Mold Star 15 SLOW (Smooth-On Inc., Macungie, PA, USA) was mixed in a 1A:1B mass ratio, per manufacturer's recommendation, and poured into the negative mold. Care was taken to degas and remove bubbles when mixing the Mold Star 15 SLOW resins. The silicone positive mold was allowed to cure for 48 hours at room temperature, after which it was removed from the aluminum negative mold. The resulting positive silicone mold was cleaned with acetone to remove debris, organic matter, or residual resin from previous batches, and sequentially used to cast WC-55 polyurethane disks in a 1A:2B mass ratio mixture. Similar to preparation of the Mold Star mixture, care was taken to degas the mixture before pouring into the positive mold to avoid formation of air bubbles in the pouring process. Once poured, the mold was left at room temperature for 24 hours to cure, after which the mold was placed in an oven at 50 $^oC$ for 24 hours to ensure a more complete chemical reaction. Photoelastic particles were then de-molded and checked for quality.

\subsection{Photoelastic Calibration Experiment}
\label{sec: Photoelastic Calibration Experiment}

The stress-optic coefficient, $C$, is a material property quantifying the photoelastic response of a material. Typically, this parameter is found by performing diametric compression testing \cite{pegspaper,photoelastic_wiki_net} to count the number of fringes with increasing normal load, measured with an external load cell. In this study, we used a custom test frame to apply and measure diametric loading. A particle is compressed between a micrometer-actuated translating stage and a fixed load cell, see figure \ref{fig: fringe fitting and test setup}g. We used an Ametek DFS3 series force gauge with AFM-010 external load cell to measure force magnitudes at the lower contact point. Images of the particle were recorded using a $16 \ MP$ monochrome ISVI IC-X16RZ-CX1 digital camera. The test setup was illuminated using a Phlox HSC white LED backlight and a polariscope was created using Edmund Optics left- and right-handed circular polarizing polymer films. Since the stress-optic coefficient is a wavelength dependent property, a HOYA HMC red lens filter was mounted to the camera lens to ensure monochromatic light.
\\
\\
Once diametrically loaded particles were imaged through the polariscope, the experimental fringe pattern is fitted to its theoretical fringe pattern via least-squares non-linear optimization. We modified the open-source photoelastic grain solver (PEGS) code \cite{pegspaper} to fix the force magnitude, force direction, and contact points and solved for the stress-optic coefficient via an inverse problem, see \cite{PEGS_Correction_McMillan_2025}. Figure \ref{fig: fringe fitting and test setup} illustrates this process. Here, $F_\sigma$ is related to the stress-optic coefficient by:
\begin{gather}
    \label{eq: fsigma definition}
    F_\sigma = \frac{\lambda}{Ch}
\end{gather}
where $\lambda$ is the wavelength of light, $C$ is the stress-optic coefficient, and $h$ is the particle thickness. The theoretical fringe data is then computed using the stress-optic law
\begin{gather}
    \label{eq: stress-optic law}
    I(x,y) = I_o sin^2 \left[\dfrac{\pi \left( \sigma_1 - \sigma_2 \right)}{F_\sigma}\right]
\end{gather}
where $I(x,y)$ is the synthetic image fringe data, $I_o$ is the initial image intensity, and $\sigma_1$ and $\sigma_2$ are the first and second principal stresses respectively. A least-squares fit of the experimental image data to the synthetic image data gives $F_\sigma$ and hence the stress-optic coefficient. As shown in figure \ref{fig: fringe fitting and test setup}, a minimum number of fringes is needed to accurately resolve $F_\sigma$ and reduce errors. Here, we have found that this minimum number is approximately one fringe, which typically occurs for forces over $2 \ N$ with WC-55, resulting in a constant material property $F_\sigma$. 
Therefore, we select data for which the applied normal force is $2 \ N$ or larger to compute the mean and standard deviation of $F_\sigma$.

\subsection{Swelling/De-Swelling Test Setup}
\label{sec: swelling/de-swelling test setup}

Fluid swelling into the cast polyurethane matrix is achieved via direct immersion in water. We define the mass-based swelling capacity as
\begin{gather}
    \label{eq: swelling capacity mass}
    SC_{mass} = 100 \times \left( \frac{m_p - m_o}{m_o} \right)
\end{gather}
where $m_p$ is the instantaneous mass of the particle and $m_o$ is the initial dry mass of the particle before swelling. Particles were placed in a beaker of water with the top sealed with parafilm to prevent evaporation. At discrete times throughout the swelling experiments, the mass of each particle was measured using an AX523 Adventurer precision balance. Particles were removed from the beaker of water, air-dried using compressed air for 10 seconds to remove water on the surface, and then weighed for each swelling time interval. After measuring the mass, digital calipers were used to measure the dimensions of the sample. Next, the stress-optic coefficient was measured for each dry particle, as discussed in section \ref{sec: Photoelastic Calibration Experiment}.
\\
\\
Throughout the swelling process, the particle's mechanical and optical properties change due to the influx of water. After a critical swelling time, measured in weeks, a particle reaches its maximum swelling extent where water no longer diffuses into the polyurethane matrix. At this point, we annealed the particles in an oven at $70 \ ^{\circ}C$ and measured the stress-optic coefficient over time to assess their ability to de-swell.

\subsection{Rheological Analysis}
\label{sec: rheological analysis}

Consider two photoelastic particles colliding together. The impact force induces a stress field within the material which results in a strain field. In the case of a linear elastic solid \cite{theory_of_elasticity}, the stress, $\sigma$, is proportional to the strain, $\epsilon$, by Hooke's law:
\begin{gather}
    \label{eq: linear elastic solid}
    \sigma = E \epsilon,
\end{gather}
where $E$ is the elastic modulus, which is a measure of the stress response of the material under the deformation. For small strains in elastic solids, this response is instantaneous, and upon removal of the force, the material immediately returns to its original shape. However, photoelastic materials such as polyurethanes are rarely linear elastic solids. Instead, both the deformation upon loading and subsequent relaxation upon the cessation of force follow a time-dependent behavior. This phenomenon observed in most materials is termed \textit{viscoelasticity} \cite{viscoelastic_polymer_book_Ferry}. Here, the time-dependent stress is related to the applied strain. Under shear,
\begin{gather}
    \label{eq: linear viscoelastic solid}
    \sigma_s(t) = G(t) \epsilon_s
\end{gather}
where $G(t)$ is the time-dependent shear stress relaxation modulus of the material; and, for tensile, 
\begin{gather}
    \label{eq: linear viscoelastic solid tensile}
    \sigma_t(t) = E(t) \epsilon_t
\end{gather}
where $E(t)$ is the time-dependent tensile stress relaxation modulus of the material. The viscoelastic behavior of our soft granular particles plays an important role in particle-particle collisions. Particles that exhibit higher dampening and a longer stress relaxation time result in a slower relaxation of the photoelastic response \cite{Fossil_Photoelasticity_McMillan2026}. In photoelastic analog experiments of high-speed flows, such as avalanches \cite{Avalanche_Amalia_PhysRevE.100.012902}, the rheology (specifically stress relaxation time) sets the minimum resolvable experimental time scale. Therefore, it is important to quantify the rheology of \textit{wet} photoelastic particles to inform use in photoelastic suspension experiments, such as debris flows or mudslides.
\\
\\
In this study, we analyze the \textit{linear} viscoelastic regime of our photoelastic material to understand how intrinsic material properties change due to swelling. To characterize the rheology, dynamic mechanical analysis (DMA) and small-angle oscillatory shear (SAOS) experiments were performed on the cured photoelastic resin using an Anton Paar MCR 702e rheometer to get the tensile moduli ($E'(\omega)$ and $E''(\omega)$) and dynamic moduli ($G'(\omega)$ and $G''(\omega)$) data respectively. In our paper, the relaxation modulus $G(t)$ was calculated from dynamic moduli data, $G'(\omega)$ and $G''(\omega)$ referring to storage and loss modulus respectively. For SAOS, the discrete relaxation spectra are calculated as:
\begin{align}
    \label{eq: G'}
    G'(\omega) &= G_e + \sum_{i=1}^N g_i \frac{(\omega \tau_i)^2}{1 + (\omega \tau_i)^2}
\end{align}
\begin{align}
    \label{eq: G''}
    G''(\omega) &= \sum_{i=1}^N g_i \frac{\omega \tau_i}{1 + (\omega \tau_i)^2}
\end{align}
where $G'(\omega)$ is the storage modulus, $G''(\omega)$ is the loss modulus, $\omega$ is the angular frequency, $g_i$ represents the $i^{\text{th}}$ the relaxation mode having relaxation time $\tau_i$ \cite{Baumgaertel1989}. $G_e$ is the equilibrium modulus, which has a finite value for solids but zero for liquids. A similar analysis can be done for DMA to get the tensile moduli.
\\
\\
The relaxation modulus can then be calculated from the $N$ modes of the discrete relaxation spectrum as
\begin{gather}
    \label{eq: G(t)}
    G(t) = G_e + \sum_{i=1}^N g_i e^{-t/\tau_i}.
\end{gather}
The conversion from dynamic moduli data to relaxation modulus was done using the ReSpect v3 software package \cite{matlab_respect}. The relaxation modulus decays from an initial value, $G(t=0) = G_0$, to the equilibrium modulus, $G_e$. The shear modulus, $G_\infty$, is defined as
\begin{gather}
    \label{eq: shear modulus}
    G_\infty = \lim_{t \to \infty} G(t) = G_e
\end{gather}
in which the shear modulus and Young's modulus are related by
\begin{gather}
    \label{eq: elastic material}
    G_\infty = \frac{E}{2(1 + \nu)}
\end{gather}
where $E$ is the Young's modulus and $\nu$ is the Poisson's ratio \cite{theory_of_elasticity}.  
\\
\\
In this study, we found the linear viscoelastic range for WC-55 by first performing an amplitude sweep and then applying an oscillatory sweep across angular frequencies of interest. Oscillatory shear testing was performed using casted disks of $D = 8 \ mm$ and $H = 1.5 \ mm$ under an axial load of $25 \ N$ to prevent slipping at $\gamma = 0.025\%$ for $\omega = 0.5-600 \ rad/s$. Dynamic mechanical analysis (DMA) was performed using casted bars of $L = 25 \ mm$, $W = 8 \ mm$, and $H = 3 \ mm$ at $\gamma = 0.2\%$ for $\omega = 1.25 - 62.83 \ rad/s$. We performed both SAOS and DMA testing for water-immersed samples following the procedure set forth in section \ref{sec: swelling/de-swelling test setup}.


\section{Photoelastic Response to Hydration}

\begin{figure*}[t!]
    \centering
    \includegraphics[width=\linewidth]{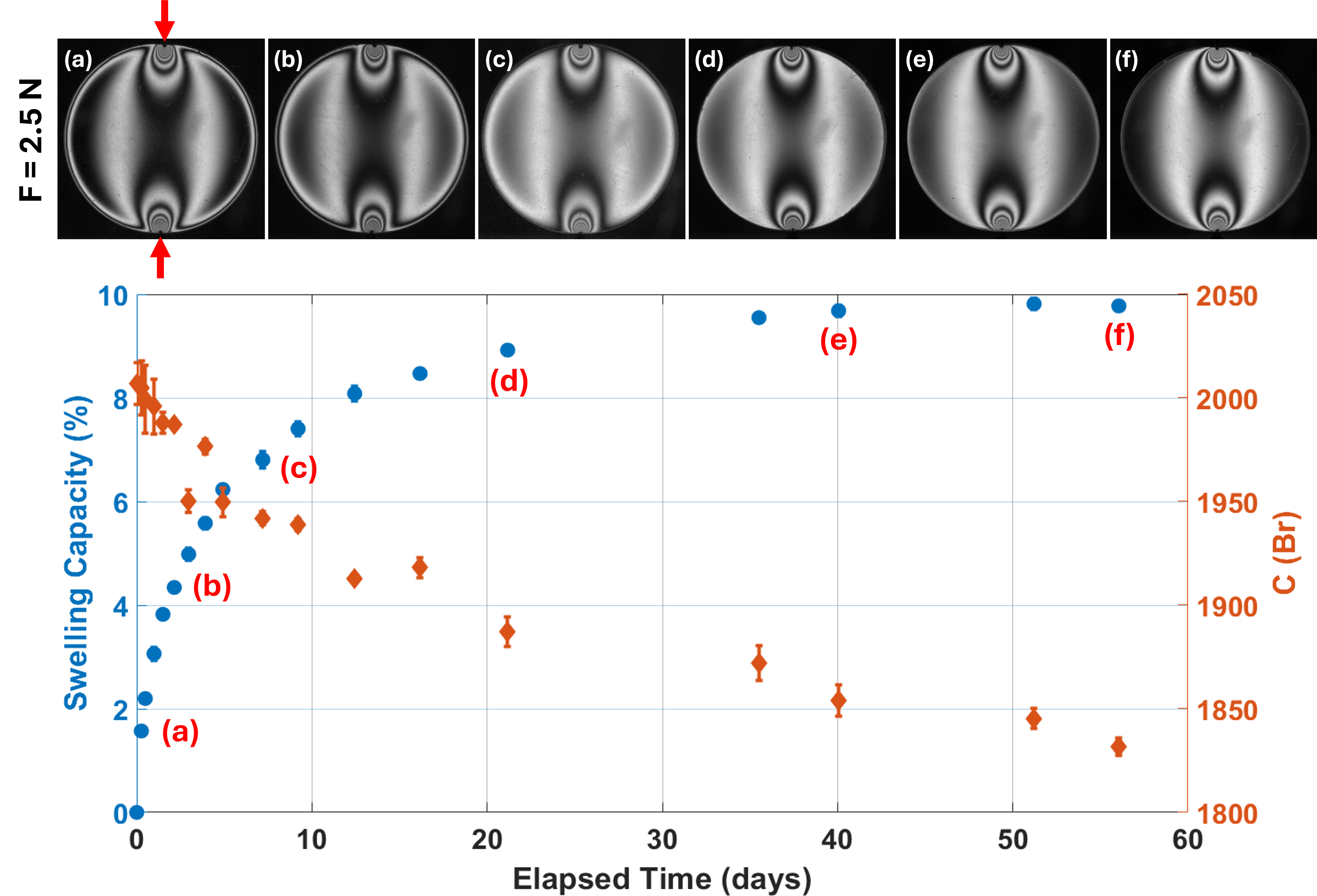}
    \caption{Effect of Swelling on the Stress-Optic Coefficient. Illustrates the inverse relationship between the swelling capacity and stress-optic coefficient, $C$. As the swelling capacity increases to its maximum at approximately $10\%$ near 55 days, the stress-optic coefficient declines by approximately $9\%$. (a-f) Show photoelastic image data for an applied diametric load of $F = 2.5 \ N$ at increasing swelling times. 3 sample replicates were performed for each data point using three water-immersed WC-55 photoelastic particles ($D = 22 \ mm$ and $H = 6 \ mm$).}
    \label{fig: photoelastic swelling data}
\end{figure*}

\subsection{Stress-Optic Coefficient Decay}

\begin{figure}[t!]
    \centering
    \includegraphics[width=\linewidth]{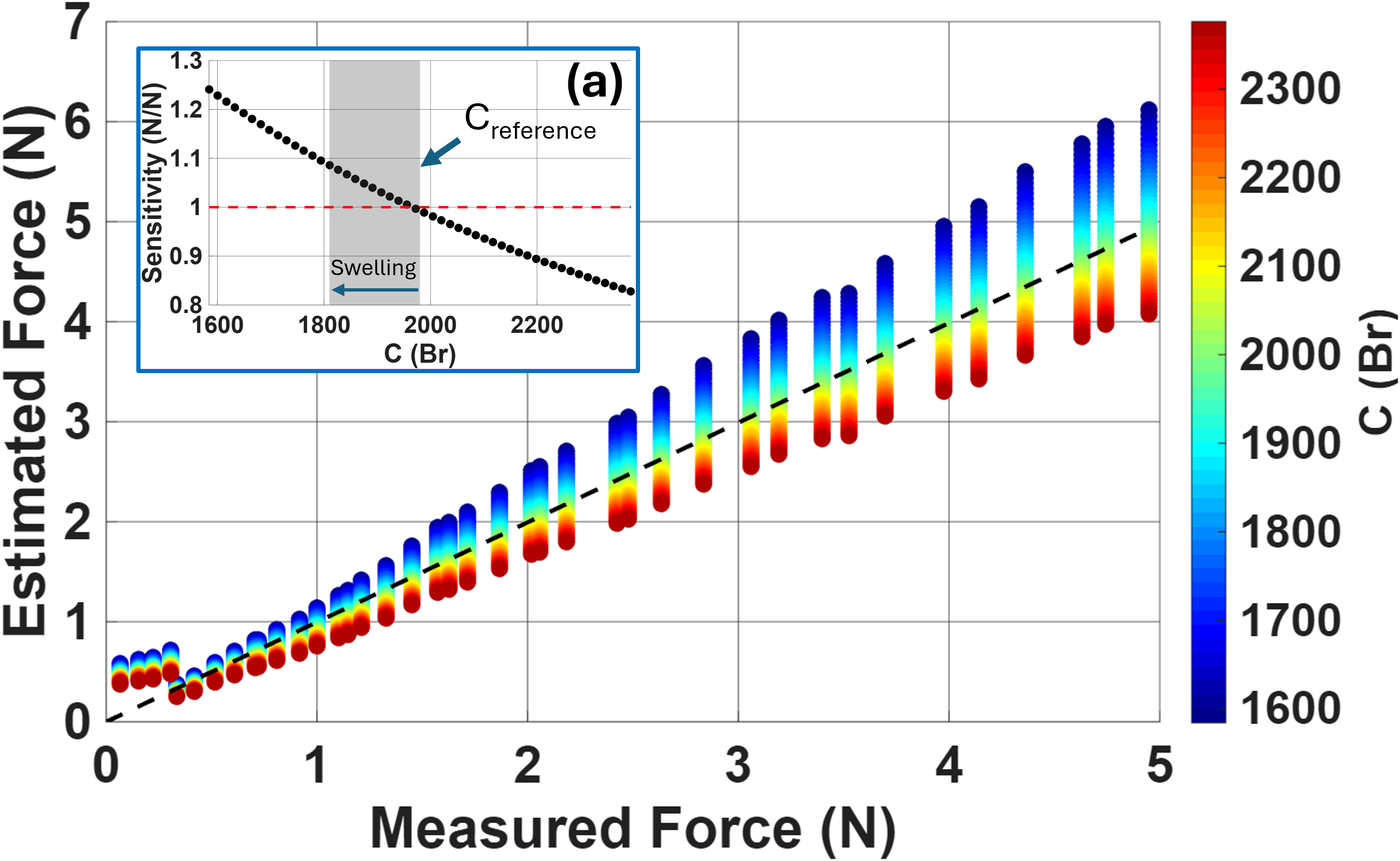}
    \caption{Stress-Optic Coefficient Sensitivity on PEGS Force Estimation. Using the dataset from figure \ref{fig: fringe fitting and test setup}, the stress-optic coefficient is varied 20$\%$ from its reference of $C = 1980 \ Br$ and the PEGS code estimated force is compared to the experimental measured force. A perfect PEGS estimation would give a slope, which we define as the sensitivity, of 1. (a) Illustrates how the sensitivity changes as a function of the stress-optic coefficient, denoting the measured amount swelling changes the stress-optic coefficient by shading. The dashed line denotes a sensitivity of 1.}
    \label{fig: C sensitivity}
\end{figure}

By immersing photoelastic experiments in fluids, we are able to mimic a range of geophysical suspensions, such as mudslides and debris flows. Here, we address the extent to which the stress-optic coefficient changes as a function of swelling capacity. Three WC-55 photoelastic particles ($D = 22 \ mm$ and $H = 6 \ mm$) were immersed in water and the swelling capacity and stress-optic coefficient were measured at discrete time intervals over approximately 2 months. Figure \ref{fig: photoelastic swelling data} illustrates how the swelling capacity and stress-optic coefficient, $C$, change throughout hydration. Initially, the rate of swelling is high, but then water begins to saturate the polyurethane matrix reducing the rate of swelling to near 0 at around 55 days. For our WC-55 cylindrical particles, the maximum swelling capacity is approximately $10\%$ after 55 days. We observe that the stress-optic coefficient shows an inverse relationship with the swelling capacity. Dry WC-55 photoelastic particles started with a stress-optic coefficient of approximately $2006 \ Br$ which then declined to $1831 \ Br$ after 55 days, a $9\%$ decrease. A reduction in the stress-optic coefficient implies lower sensitivity for resolving forces on swollen particles since less fringes per given load are present, see the image series in figure \ref{fig: photoelastic swelling data}a-f. We note that initial water swelling into the polyurethane matrix creates a water-induced stress halo observable in figure \ref{fig: photoelastic swelling data}a-c, which is discussed in detail in the next section. 
\\
\\
Quantitative photoelasticity assumes that the stress-optic coefficient is constant. Yet, for water-immersed particles, we find a function of time. In the classical use of the PEGS code, the contact points, stress-optic coefficient, and dimensions of a particle are known and fringe inversion provides the vector forces acting at all contact points. Figure \ref{fig: C sensitivity} presents a sensitivity analysis quantifying how changes in the stress-optic coefficient effect the PEGS code force estimation using the dataset from figure \ref{fig: fringe fitting and test setup}. Ideally, the estimated force would exactly match the measured force. In this case, a plot of estimated force versus measured force would have a slope of 1, shown as the solid dashed line. We denote the slope of this line as the \textit{sensitivity}. A sensitivity of 1.10 would mean that the PEGS code overestimates the force by 10$\%$ of its actual value. Shown in figure \ref{fig: C sensitivity}, as the stress-optic coefficient varies from its reference value of $C_{ref} = 1980 \ Br$, the force prediction goes from overestimating ($C < C_{ref}$) to underestimating ($C > C_{ref}$). The inset in figure \ref{fig: C sensitivity} plots the sensitivity as a function of the stress-optic coefficient where the shaded portion represents the local change swelling has on the stress-optic coefficient. Thus, a $9\%$ change in the stress-optic coefficient due to swelling causes a $9\%$ increase in the sensitivity, meaning an overestimate of the contact forces.

\subsection{Water-Induced Stress Halo}

\begin{figure*}[t!]
    \centering
    \includegraphics[width=\linewidth]{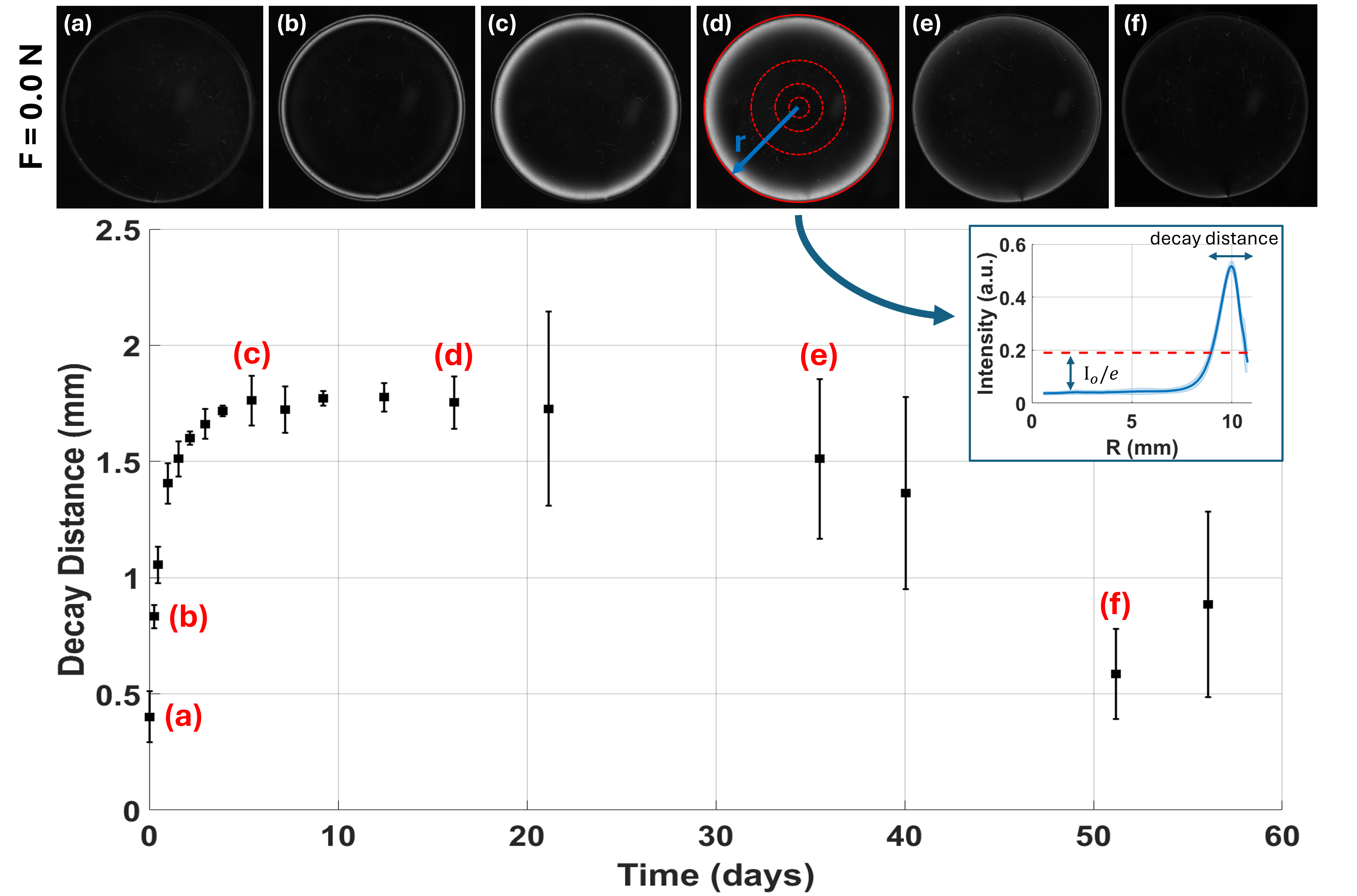}
    \caption{Water-Induced Stress Halo Phenomenon. (a-f) Show diffusion of water into the polyurethane matrix which creates internal stress around the particle's edge that is observable in the photoelastic signal. The inset illustrates the definition of the decay distance defined as the distance from the edge of the particle towards the center where the photoelastic signal intensity reaches $I_o/e$. $N = 3$ sample replicates were performed for each data point using three water-immersed WC-55 photoelastic particles ($D = 22 \ mm$ and $H = 6 \ mm$).}
    \label{fig: water diffusion halo}
\end{figure*}

\begin{figure}[t!]
    \centering
    \includegraphics[width=\linewidth]{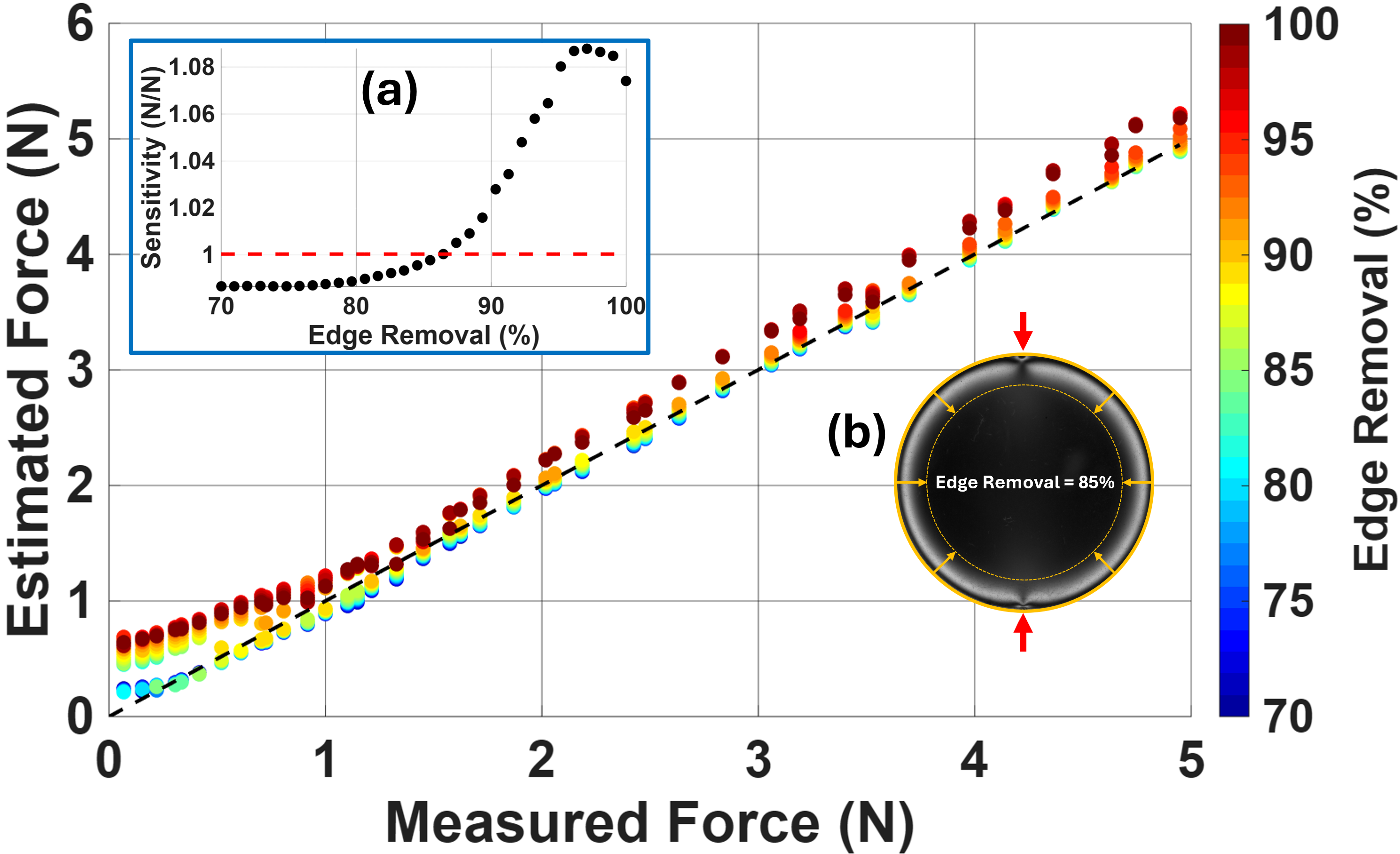}
    \caption{Edge Removal Sensitivity on PEGS Force Estimation. Using the dataset from figure \ref{fig: fringe fitting and test setup} with a fixed stress-optic coefficient of $C = 1980 \ Br$, the edge removal percentage is varied from $70-100\%$ and the PEGS code estimated force is compared to the experimental measured force. A perfect PEGS estimation would give a slope, which we define as the sensitivity, of 1. (a) Illustrates how the sensitivity changes as a function of the edge removal percentage and panel (b) shows how an edge removal of $85\%$ effectively removes the water-induced stress halo from the fringe inversion.}
    \label{fig: edge removal sensitivity}
\end{figure}

After WC-55 particles are immersed in water for a few hours, an observable photoelastic signal begins to appear around the edge of the particles, see figure \ref{fig: water diffusion halo}a-f. We term this phenomenon \textit{water-induced stress halo}. Mechanistically, water diffuses into the polyurethane matrix creating internal stress in the particle which we observe as a photoelastic halo around the particle's edge. As the swelling time increases, the radius of the halo increases while the signal intensity decreases until the decay distance, the distance from the edge of the particle towards the center where the photoelastic signal intensity reaches $I_o/e$, peaks at approximately 8 days with a decay distance of approximately $1.75 \ mm$. As the swelling time further increases, the water permeates the entire diameter of the particle and the decay distance decreases to near its initial starting value. We note that the initial starting value is not at a decay distance of 0 since the edge effect around the photoelastic particles creates a small photoelastic signal.
\\
\\
Practically, the water-induced stress halo phenomenon reduces the accuracy of PEGS fringe inversion for low force values since the optimization solver sees photoelastic signal around the edge of the particle instead of a nearly dark particle. This effect can be compensated for by increasing the ``edge removal percentage'' in the PEGS code which filters the amount of the experimental image being fitted in the PEGS fringe inversion, effectively removing the halo during the fitting process as shown in figure \ref{fig: edge removal sensitivity}b. However, for small forces, the experimental fringe pattern at the edge of the particle is often important to properly resolve the force. Therefore, a larger edge removal percentage means that the sensitivity to lower force magnitudes decreases. Figure \ref{fig: edge removal sensitivity} plots the PEGS estimated force versus the experimental measured force for edge removal values ranging from $70-100\%$. An edge removal of 1.00 or 100$\%$ means the entire image is used in the fringe inversion. Here, the reference stress-optic coefficient is fixed at $C = 1980 \ Br$ using the dataset from figure \ref{fig: fringe fitting and test setup}. We found that an edge removal of 0.85 eliminates the bias from the water-induced stress halo but limits the lower resolvable force bound to $0.30 \ N$. In this study, an edge removal of 0.85 is used for every PEGS analysis showing water-induced stress halos. Figure \ref{fig: edge removal sensitivity}a shows the sensitivity as a function of edge removal percentage in which increasing the edge removal percentage causes the PEGS code to overestimate forces.
    
\subsection{Water Removal via Annealing}

\begin{figure*}[t!]
    \centering
    \includegraphics[width=\linewidth]{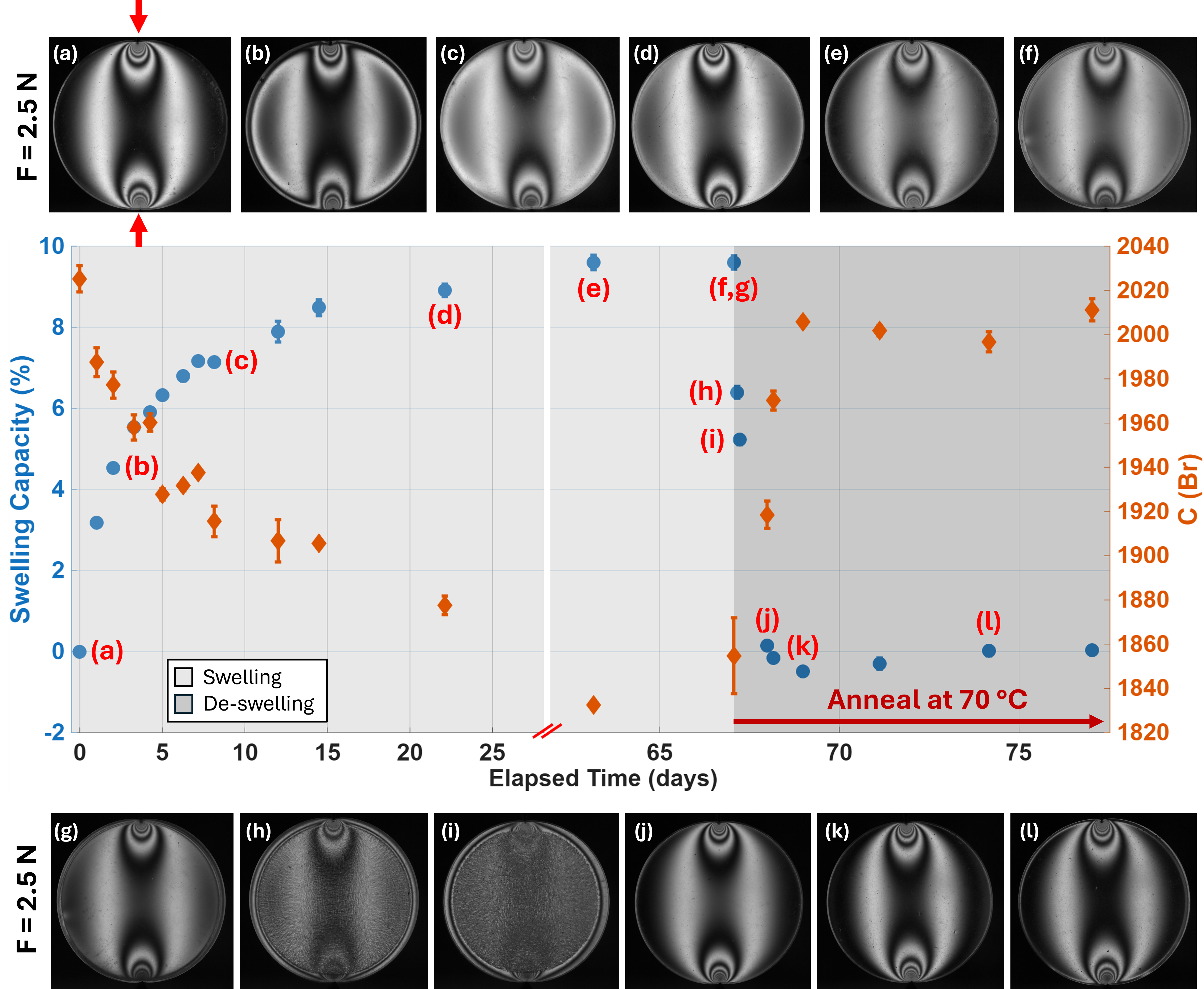}
    \caption{De-Swelling of Water-Immersed Photoelastic Particles. Illustrates the recovery of the stress-optic coefficient as water is removed from the particles at $70 \ ^{\circ}C$. After 10 days of annealing, the swelling capacity returns to near 0$\%$ and the stress-optic coefficient nearly completely returns to its initial dry state. Panels (a-f) depict photoelastic images of particles under diametric loading at $2.5 \ N$ as swelling time increases. Panels (g-l) depict photoelastic images of particles under diametric loading at $2.5 \ N$ during de-swelling at $70 \ ^{\circ}C$. $N = 3$ sample replicates were performed for each data point using three water-immersed WC-55 photoelastic particles ($D = 22 \ mm$ and $H = 6 \ mm$).}
    \label{fig: de-swelling figure}
\end{figure*}

In a typical granular flow experiment, thousands of photoelastic particles are used which take many months to fabricate. A permanent state change due to swelling would mean that experiments requiring dry initial conditions would need new particles fabricated for every experimental run, quickly becoming intractable. Therefore, there is a need to ``de-swell'' photoelastic particles where swollen particles can be returned to near dry initial conditions. 
\\
\\
Here, we swell three WC-55 particles ($D = 22 \ mm$ and $H = 6 \ mm$) in water over approximately 2 months and measure the swelling capacity and stress-optic coefficient according to section \ref{sec: swelling/de-swelling test setup}. Once fully saturated with water, the swollen particles were annealed in an oven at $70 \ ^{\circ}C$ to remove the water. Swelling capacity and the stress-optic coefficient were measured at discrete time intervals over approximately 10 days. Figure \ref{fig: de-swelling figure} show both swelling and de-swelling of WC-55 particles. Panels (a-f) depict photoelastic images of particles under diametric loading of $2.5 \ N$ as swelling time increases. Panels (g-l) depict photoelastic images of particles under diametric loading of $2.5 \ N$ during de-swelling at $70 \ ^{\circ}C$. De-swelling at $70 \ ^{\circ}C$ rapidly removed water from the WC-55 particles in approximately 2 days upon which near full recovery of the stress-optic coefficient occurred after 10 days at a swelling capacity of $0\%$. We note that rapid evaporation of water from the polyurethane matrix initially caused optical interference, see figure \ref{fig: de-swelling figure}h-i, after which the photoelastic signal returned to normal. 

\section{Rheological Response to Hydration}

\begin{figure*}[t!]
    \centering
    \includegraphics[width=\linewidth]{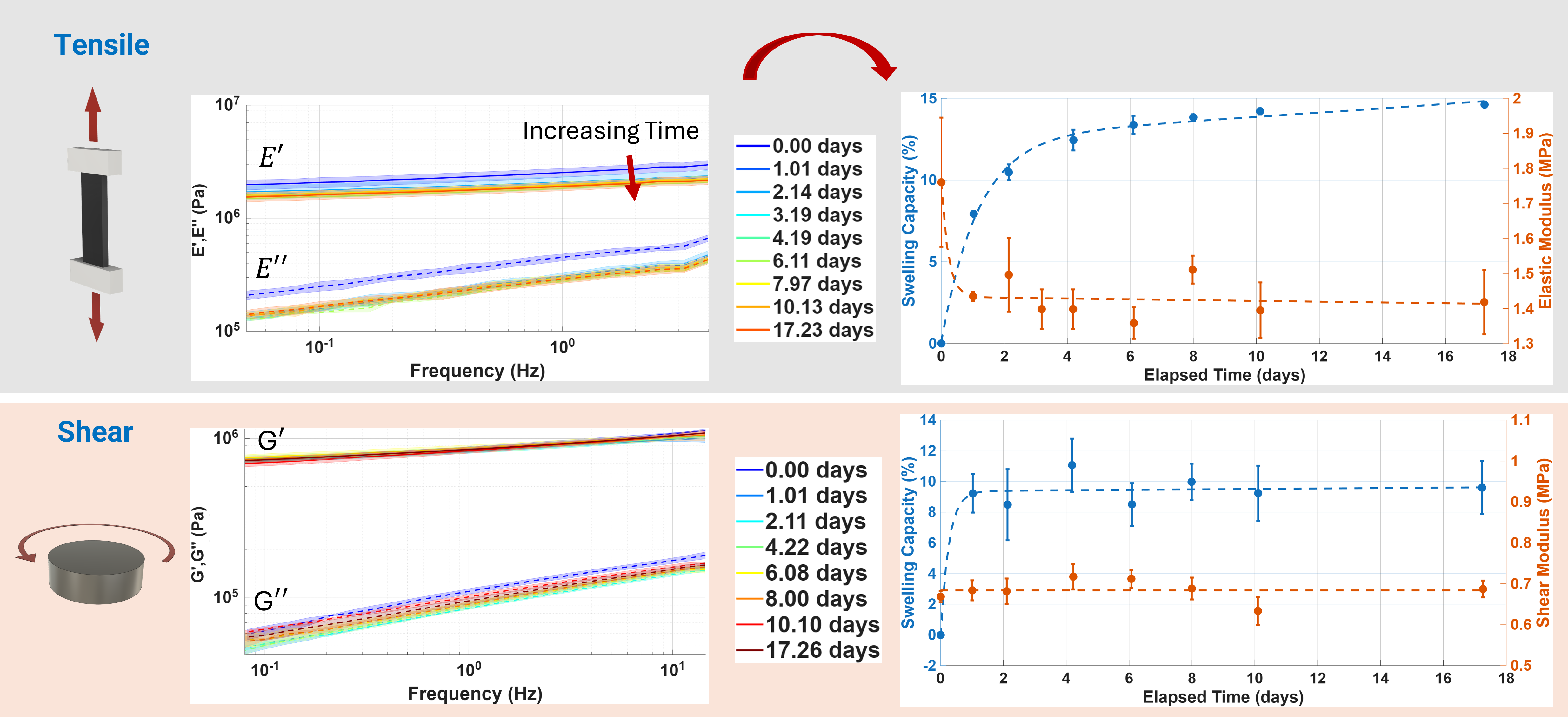}
    \caption{Mechanical Characterization of Hydrated WC-55 Particles. Dynamic mechanical analysis and small-angle oscillatory shear testing were performed to characterize the tensile and shear storage and loss moduli as a function of swelling time, see section \ref{sec: rheological analysis} for more details. We note that $G_\infty \neq 3E$ as theoretically expected. We hypothesize that sample slip occurs during shear testing and thus we rely on tensile rheological testing for more accurate mechanical properties measurement. A decrease in the tensile storage and loss moduli as a function of swelling time indicates that swelling causes the polymer to slightly soften which can be seen by the reduction in the elastic modulus. $N = 3$ sample replicates were performed for each data point using three water-immersed WC-55 photoelastic disks ($D = 8 \ mm$ and $H = 1.5 \ mm$) and rods ($L = 25 \ mm$, $W = 8 \ mm$, $H = 3 \ mm$) for shear and tensile testing respectively.}
    \label{fig: DMA results swelling}
\end{figure*}

\begin{figure}[t!]
    \centering
    \includegraphics[width=\linewidth]{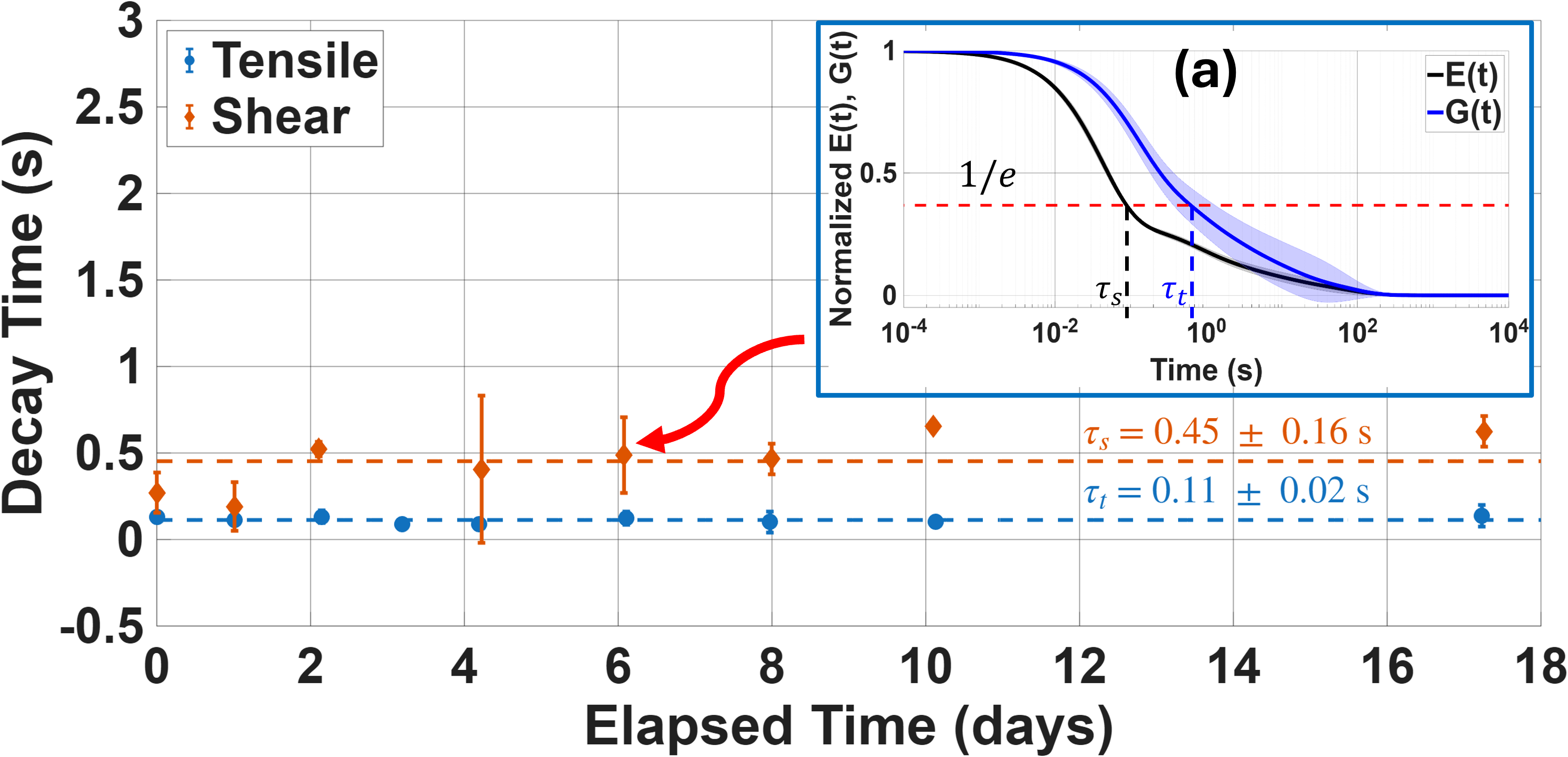}
    \caption{Effect of Swelling on the Decay Time. Illustrates how the decay time changes as a function of swelling time. (a) The decay time is defined as the time required for the normalized time-dependent stress relaxation modulus to decay to a value of 1/e, denoted by the dashed line. The average decay times were $\tau$ = 0.45 $\pm$ 0.16 $s$ and $\tau$ = 0.11 $\pm$ 0.02 $s$ for shear and tensile testing respectively. N = 3 sample replicates were performed for each data point.}
    \label{fig: decay time vs swelling}
\end{figure}

The previous section discusses the relationship between the optical properties (stress-optic coefficient) of WC-55 particles and its swelling capacity. Here, we characterize the relationship between the \textit{mechanical properties} and its swelling capacity. Small-angle oscillatory shear and dynamic mechanical testing were performed on hydrated test samples for approximately 3 weeks, as detailed in section \ref{sec: rheological analysis}. Figure \ref{fig: DMA results swelling} displays the tensile and shear storage and loss moduli as a function of swelling time. As the swelling time increases, both the tensile storage and loss moduli decrease in magnitude reaching a steady-state value around 10 days. The elastic modulus is plotted showcasing that WC-55 photoelastic particles soften while hydrating. We note that the elastic modulus changes from its dry initial state of approximately $1.8 \ MPa$ to $1.4 \ MPa$, a $22\%$ decrease. For the small-angle oscillatory shear testing, we note that the storage modulus remains approximately constant while the loss modulus decreases as a function of swelling time yielding a shear modulus that is constant over swelling time. From a theoretical rheology perspective, the shear modulus is directly related to the elastic modulus by equation \ref{eq: shear modulus}. Assuming WC-55 to be isotropic and thus a Poisson's ratio of 0.50, equation \ref{eq: shear modulus} reduces to $G_\infty = 3E$. This does not hold true in the case of our data. We hypothesize that the small-angle oscillatory shear testing is confounded by slip due syneresis of water from the sample. Even though care was taken to dry the surface of the sample thoroughly, expulsion of liquid from an infused network is unavoidable \cite{syneresis_water_coming_out_of_polymers}. Invariably, the water film causes the sample to slip which results in an incorrect lower moduli data, despite the application of a large normal force on the sample. As such, we rely on tensile rheological testing since samples can be clamped reducing the concern about slip.
\\
\\
As water diffuses into the polyurethane matrix, the polymer network swells and the elastic modulus decreases. To characterize the time-dependent dissipative behavior as a function of swelling time, we measure the decay time, which we define as the time required for the normalized time-dependent stress relaxation modulus to decay to a value of 1/e. As shown in figure \ref{fig: decay time vs swelling}, the average decay times were $\tau$ = 0.45 $\pm$ 0.16 $s$ and $\tau$ = 0.11 $\pm$ 0.02 $s$ for shear and tensile testing respectively with N = 3 sample replicates performed for each data point. Figure \ref{fig: decay time vs swelling}a illustrates the procedure for finding the decay time from the normalized time-dependent stress relaxation modulus. A constant decay time as a function of elapsed swelling time indicates that the diffusion of water into the polyurethane matrix does not contribute to the relaxation time of the material. Similar to above, a theoretical rheology perspective would imply that the decay time from tensile and shear testing would be equivalent. However, we find that the two decay times are different, further evidence for slipping under shear testing for swollen samples.


\section{Conclusions}

Historically, quantitative photoelasticity has provided insights into a wide range of dry granular phenomena. Yet, many important granular flows in nature are wet (e.g. mudslides and debris flows) where hydrodynamic stress can play a non-trivial role. Extending quantitative photoelasticity to wet granular flows has great potential to increase our understanding and ability to predict wet geophysical processes and flows. In this work, a comprehensive characterization of how hydration-induced swelling modifies the rheological and optical properties of water-clear 55 (WC-55) photoelastic material is performed. Specifically, it was found that cylindrical WC-55 particles ($D = 22 \ mm$ and $H = 6 \ mm$) fully saturate in water after 55 days with a swelling capacity of approximately 10$\%$ which causes the stress-optic coefficient to decrease by approximately 9$\%$ and the elastic modulus decreases by 22$\%$. Additionally, it was found that swelling of WC-55 particles creates internal stress around the particle's edge due to water diffusion which we termed \textit{water-induced stress halo}. It is important to filter the amount of the experimental image being fitted in the PEGS fringe inversion via the ``edge removal percentage'' to effectively remove the halo from the fringe inversion during the fitting process. 
\\
\\
We envision this work to serve as a guidebook for experimentation with water-immersed photoelastic materials and to establish a foundation for extending particle-scale force measurements to wet granular systems. Briefly, we summarize key considerations for use of photoelastic particles in immersion-based experiments below:

\begin{enumerate}
    \item Swelling induces changes in the stress-optic coefficient for photoelastic particles. Photoelastic particles should be pre-swelled such that they are fully saturated before use or select particles should be measured throughout a wet photoelastic experiment to characterize the changing stress-optic coefficient over time as particles swell.
    \item Water diffusion causes internal stress around the edge of the particles which is superimposed on the experimental fringe pattern. Particles should be pre-swelled such that they reach full saturation before use in experiments such that the water-induced stress halo is negligible.
    \item If particles are not pre-swelled, the ``edge removal percentage'' in the PEGS code should be set to remove the halo from the fringe fitting process. Note that setting a higher edge removal percentage causes a lower resolution for the minimum detectable force bound.
    \item Particles can be returned to their initial dry state by annealing out the water. For WC-55 polyurethane, swollen particles return to their dry state after 10 days in an oven at 70$^{\circ}$C.
    \item Swelling softens the mechanical properties of photoelastic materials. For WC-55 polyurethane, the material softens by 22$\%$. However, swelling has no impact on the stress relaxation decay time, measured in the linear viscoelastic regime.
\end{enumerate}

\section*{Conflicts of Interest}

There are no conflicts of interest to declare.

\section*{Data Availability}

Data is available at reasonable request to the corresponding author.

\section*{Acknowledgments}

This work has been supported by an Experimental Physics Initiative Investigator award from the Gordon and Betty Moore Foundation, grant DOI
10.37807/gbmf12236, and by startup funds to Nathalie M. Vriend provided by the University of Colorado Boulder.
\\
\\
The authors would like to thank Karen Daniels for shipping us samples of Vishay PhotoStress\texttrademark{} PSM-4 to test.









\bibliographystyle{unsrt}
\bibliography{RelevantReferences}

\begin{thebibliography}{10}

\bibitem{jenkins_theory_1983}
J.~T. Jenkins and S.~B. Savage.
\newblock A theory for the rapid flow of identical, smooth, nearly elastic,
  spherical particles.
\newblock {\em Journal of Fluid Mechanics}, 130:187--202, 1983.

\bibitem{GDRMiDi2004}
GDR MiDi.
\newblock On dense granular flows.
\newblock {\em The European Physical Journal E}, 14(4):341--365, 2004.

\bibitem{dauxois_confronting_2021}
T.~Dauxois, T.~Peacock, P.~Bauer, C.~P. Caulfield, C.~Cenedese, C.~Gorl\'{e},
  G.~Haller, G.~N. Ivey, P.~F. Linden, E.~Meiburg, N.~Pinardi, N.~M. Vriend,
  and A.~W. Woods.
\newblock Confronting Grand Challenges in environmental fluid mechanics.
\newblock {\em Physical Review Fluids}, 6(2):020501, 2021.

\bibitem{jaeger_physics_1992}
H.~M. Jaeger and Sidney~R. Nagel.
\newblock Physics of the Granular State.
\newblock {\em Science}, 255(5051):1523--1531, 1992.

\bibitem{Wakabayashi1950}
Takao Wakabayashi.
\newblock Photo-elastic Method for Determination of Stress in Powdered Mass.
\newblock {\em Journal of the Physical Society of Japan}, 5(5):383--385, 1950.

\bibitem{Drescher1972}
A.~Drescher and G.~de Josselin de Jong.
\newblock Photoelastic verification of a mechanical model for the flow of a
  granular material.
\newblock {\em Journal of the Mechanics and Physics of Solids}, 20(5):337--340,
  1972.

\bibitem{Jaeger1996}
Heinrich~M. Jaeger, Sidney~R. Nagel, and Robert~P. Behringer.
\newblock The Physics of Granular Materials.
\newblock {\em Physics Today}, 49(4):32--38, 1996.

\bibitem{howell_1999}
Daniel~W. Howell, R.~P. Behringer, and C.~T. Veje.
\newblock Fluctuations in granular media.
\newblock {\em Chaos: An Interdisciplinary Journal of Nonlinear Science},
  9(3):559--572, 1999.

\bibitem{behringernature}
T.~S. Majmudar and R.P. Behringer.
\newblock Contact force measurements and stress anisotropy in granular
  materials.
\newblock {\em Nature}, 435:1079--1082, 2005.

\bibitem{Bares2017}
Jonathan Bar\'{e}s, Dengming Wang, Dong Wang, Thibault Bertrand, Corey~S.
  O'Hern, and Robert~P. Behringer.
\newblock Local and global avalanches in a two-dimensional sheared granular
  medium.
\newblock {\em Physical Review E}, 96(5):052902, 2017.

\bibitem{Majmudar2007}
T.~S. Majmudar, M.~Sperl, S.~Luding, and R.~P. Behringer.
\newblock Jamming Transition in Granular Systems.
\newblock {\em Physical Review Letters}, 98(5):058001, 2007.

\bibitem{Clark2012}
Abram~H. Clark, Lou Kondic, and Robert~P. Behringer.
\newblock Particle Scale Dynamics in Granular Impact.
\newblock {\em Physical Review Letters}, 109(23):238302, 2012.

\bibitem{owens_sound_2011}
E.~T. Owens and K.~E. Daniels.
\newblock Sound propagation and force chains in granular materials.
\newblock {\em Europhysics Letters}, 94(5):54005, 2011.

\bibitem{mahabadi_impact_2017}
Nariman Mahabadi and Jaewon Jang.
\newblock The impact of fluid flow on force chains in granular media.
\newblock {\em Applied Physics Letters}, 110:041907, 2017.

\bibitem{Tang2018}
Zhu Tang, Theodore~A. Brzinski, Michael Shearer, and Karen~E. Daniels.
\newblock Nonlocal rheology of dense granular flow in annular shear experiments.
\newblock {\em Soft Matter}, 14(16):3040--3048, 2018.

\bibitem{abed_zadeh_enlightening_2019}
Aghil Abed~Zadeh, Jonathan Bar\'{e}s, Theodore~A. Brzinski, Karen~E. Daniels,
  Joshua Dijksman, Nicolas Docquier, Henry~O. Everitt, Jonathan~E. Kollmer,
  Olivier Lantsoght, Dong Wang, Marcel Workamp, Yiqiu Zhao, and Hu Zheng.
\newblock Enlightening force chains: a review of photoelasticimetry in granular
  matter.
\newblock {\em Granular Matter}, 21(4):83, 2019.

\bibitem{pegspaper}
Karen~E. Daniels, Jonathan~E. Kollmer, and James~G. Puckett.
\newblock Photoelastic force measurements in granular materials.
\newblock {\em Review of Scientific Instruments}, 88(5):051808, 2017.

\bibitem{PEGS_Correction_McMillan_2025}
Benjamin McMillan, Stuart~B Dalziel, and Nathalie~M Vriend.
\newblock Validation and correction of photoelastic techniques for frictional
  granular systems.
\newblock {\em Measurement Science and Technology}, 36(5):055212, 2025.

\bibitem{Fossil_Photoelasticity_McMillan2026}
Benjamin McMillan, Stuart~B. Dalziel, and Nathalie~M. Vriend.
\newblock Dynamic response, viscoelastic relaxation and boundary deformation in
  granular photoelasticity.
\newblock {\em Granular Matter}, 28(2):32, 2026.

\bibitem{thomas_2019a}
A.~L. Thomas and N.~M. Vriend.
\newblock Photoelastic study of dense granular free-surface flows.
\newblock {\em Phys. Rev. E}, 100(1):012902, 2019.

\bibitem{thomas_2019b}
A.~L. Thomas, Zhu Tang, Karen~E. Daniels, and N.~M. Vriend.
\newblock Force fluctuations at the transition from quasi-static to inertial
  granular flow.
\newblock {\em Soft Matter}, 15(42):8532--8542, 2019.

\bibitem{poon_2023}
Rebecca~N Poon, Amalia~L Thomas, and Nathalie~M Vriend.
\newblock Microscopic origin of granular fluidity: An experimental
  investigation.
\newblock {\em Physical Review E}, 108(6):064902, 2023.

\bibitem{Fazelpour2023}
Farnaz Fazelpour and Karen~E. Daniels.
\newblock Controlling rheology \textit{via} boundary conditions in dense
  granular flows.
\newblock {\em Soft Matter}, 19(12):2168--2175, 2023.

\bibitem{guazzelli_rheology_2018}
\'{E}lisabeth Guazzelli and Olivier Pouliquen.
\newblock Rheology of dense granular suspensions.
\newblock {\em Journal of Fluid Mechanics}, 852:P1, 2018.

\bibitem{stabile_clearflex50_reference}
Christopher~J. Stabile and Kevin~T. Turner.
\newblock High Strength and Dynamically Tunable Adhesion Enabled by Composite
  Micropillar Arrays Fabricated via Solvent-Assisted Molding.
\newblock {\em Advanced Materials Technologies}, 10(7):2401588, 2025.

\bibitem{cramer2014quantification}
K.~Elliott Cramer, Maurice Hayward, and William~T. Yost.
\newblock Quantification of Residual Stress from Photonic Signatures of Fused
  Silica.
\newblock In {\em AIP Conference Proceedings}, volume 1581, pages 1679--1686.
  AIP Publishing, 2014.

\bibitem{vishay_photostress_coatings_2015}
Vishay Precision Group, Micro-Measurements.
\newblock {\em PhotoStress\textsuperscript{\textregistered} Coatings}.
\newblock Document No.\ 11222, August 2015.
\newblock Revision: 06-Aug-2015.
\newblock \url{https://vishay-straingauge.com/pdf/photostress/11222BPhoto.pdf}.

\bibitem{smoothon_clearflex5095_tb}
Smooth-On, Inc.
\newblock {\em Clear Flex\textsuperscript{\textregistered} Series Technical
  Bulletin}.
\newblock 2025.
\newblock Technical bulletin for Clear Flex 50 and Clear Flex 95 water-clear
  urethane rubbers.
\newblock \url{https://www.smooth-on.com/tb/files/CLEAR_FLEX_50_95_TB.pdf}.

\bibitem{bjb_wc55am_2025}
BJB Enterprises, Inc.
\newblock {\em WC-55AM A/B Water Clear Shore 55 A Polyurethane Elastomer}.
\newblock February 2025.
\newblock Technical data sheet, dated 2025-02-20.
\newblock \url{https://bjbmaterials.com/media/wysiwyg/pdfs/Water-Clear-Shore-A/WC-55AM.pdf}.

\bibitem{Carmeen_Lee_SFR_relationship}
Carmen~L. Lee, Ephraim Bililign, Emilien Az\'{e}ma, and Karen~E. Daniels.
\newblock Particle Scale Anisotropy Controls Bulk Properties in Sheared
  Granular Materials.
\newblock {\em Phys. Rev. Lett.}, 135(10):108201, 2025.

\bibitem{wave_propagation_granular_1991_hard_disks_SHUKLA1991165}
Arun Shukla.
\newblock Dynamic photoelastic studies of wave propagation in granular media.
\newblock {\em Optics and Lasers in Engineering}, 14(3):165--184, 1991.

\bibitem{Cox2016}
Meredith Cox, Dong Wang, Jonathan Bar\'{e}s, and Robert~P. Behringer.
\newblock Self-organized magnetic particles to tune the mechanical behavior of
  a granular system.
\newblock {\em EPL (Europhysics Letters)}, 115(6):64003, 2016.

\bibitem{Bars2016ExperimentalOO}
Jonathan Bar\'{e}s, Serge Mora, Jean-Yves Delenne, and Thierry Fourcaud.
\newblock Experimental observations of root growth in a controlled photoelastic
  granular material.
\newblock {\em Epj Web of Conferences}, 140:14008, 2016.

\bibitem{photoelastic_wiki_net}
Photoelastic methods wiki (available at
  \url{https://photoelasticity.net}).

\bibitem{theory_of_elasticity}
Stephen Timoshenko and J.N. Goodier.
\newblock {\em Theory of elasticity}.
\newblock McGraw-Hill, 2nd edition.

\bibitem{viscoelastic_polymer_book_Ferry}
John~D. Ferry.
\newblock {\em Viscoelastic properties of polymers}.
\newblock John Wiley \& Sons, third edition, 1980.

\bibitem{Avalanche_Amalia_PhysRevE.100.012902}
A.~L. Thomas and N.~M. Vriend.
\newblock Photoelastic study of dense granular free-surface flows.
\newblock {\em Phys. Rev. E}, 100(1):012902, 2019.

\bibitem{Baumgaertel1989}
M.~Baumgaertel and H.~H. Winter.
\newblock Determination of discrete relaxation and retardation time spectra from
  dynamic mechanical data.
\newblock {\em Rheologica Acta}, 28(6):511--519, 1989.

\bibitem{matlab_respect}
A.~Takeh and S.~Shanbhag.
\newblock A Computer Program to Extract the Continuous and Discrete Relaxation
  Spectra from Dynamic Viscoelastic Measurements.
\newblock {\em Applied Rheology}, 23(2), 2012.

\bibitem{syneresis_water_coming_out_of_polymers}
Chihiro Urata, Gary~J. Dunderdale, Matt~W. England, and Atsushi Hozumi.
\newblock Self-lubricating organogels (SLUGs) with exceptional
  syneresis-induced anti-sticking properties against viscous emulsions and
  ices.
\newblock {\em J. Mater. Chem. A}, 3(24):12626--12630, 2015.

\end{thebibliography}


\end{document}